\documentclass{article} 
\usepackage{iclr2027_conference,times}

\usepackage{amsmath,amsfonts,bm}

\def\eqref#1{equation~\ref{#1}}

\def\1{\bm{1}}

\DeclareMathAlphabet{\mathsfit}{\encodingdefault}{\sfdefault}{m}{sl}
\SetMathAlphabet{\mathsfit}{bold}{\encodingdefault}{\sfdefault}{bx}{n}

\DeclareRobustCommand{\tool}{\textsc{VeriSpec}\xspace}
\newcommand{\parhead}[1]{\noindent\textbf{#1}}

\definecolor{hlblue}{RGB}{205,229,255}
\definecolor{hlpink}{RGB}{255,208,224}
\definecolor{hlgray}{RGB}{226,226,226}
\definecolor{hlgreen}{RGB}{209,237,214}
\newcommand{\hlc}[2][hlyellow]{{\sethlcolor{#1}\hl{#2}}}

\definecolor{caseblue}{RGB}{45,85,135}
\definecolor{casegreen}{RGB}{47,112,99}
\definecolor{casegold}{RGB}{142,97,31}
\definecolor{caseborder}{RGB}{205,211,219}
\definecolor{casefill}{RGB}{248,250,252}
\definecolor{rqfill}{HTML}{F2F7FF}
\definecolor{rqframe}{HTML}{2E5587}

\newcommand{\topicregistrydash}{\textcolor{black!40}{\rule[0.45ex]{3pt}{0.25pt}\kern1.5pt\rule[0.45ex]{3pt}{0.25pt}}\hspace{0.35em}}

\usepackage{hyperref}
\usepackage{url}
\usepackage{xcolor}

\usepackage{amsmath,amssymb,amsthm}
\usepackage{booktabs}
\usepackage{array}
\usepackage{graphicx}
\usepackage{wrapfig}
\usepackage{caption}
\usepackage{algorithm}
\usepackage{algpseudocode}
\usepackage{enumitem}
\usepackage{tabularx}
\usepackage{xspace}

\usepackage{soul}
\definecolor{hlyellow}{RGB}{255,241,153}
\sethlcolor{hlyellow}

\usepackage[most]{tcolorbox}
\newtcolorbox{rqbox}{
  colback=rqfill, colframe=rqframe, boxrule=0.5pt,
  arc=2pt, left=4pt, right=4pt, top=3pt, bottom=3pt
}

\usepackage{tikz}
\usetikzlibrary{fit,positioning,backgrounds}

\newcolumntype{Y}{>{\raggedright\arraybackslash}X}

\newtheorem{definition}{Definition}

\title{Detecting Inconsistencies in Model Specifications with LLM-as-Verifier Reasoning}

\author{%
Zichen Xie$^{1*}$, Mrigank Pawagi$^{2*}$, Lize Shao$^{1*}$, Yang Hu$^{3}$, Wenxi Wang$^{1}$ \\
$^{1}$University of Virginia \ $^{2}$University of Pennsylvania \ $^{3}$The University of Texas at Austin \\
\texttt{\{graysonxie,zgr3et,wenxiw\}@virginia.edu} \\
\texttt{mpawagi@upenn.edu}, \texttt{huyang@utexas.edu}
}

\iclrfinalcopy 
\begin{document}

\maketitle
\lhead{Preprint.}
{\renewcommand{\thefootnote}{\fnsymbol{footnote}}
\footnotetext[1]{Equal contribution.}}

\begin{abstract}
Model specifications define how large language models (LLMs) should behave, guiding alignment training, inference-time behavior, and evaluation. Yet these specifications may themselves contain defects: two individually reasonable principles may prescribe incompatible behavior when applied to the same situation, leaving no response that satisfies both. Detecting such inconsistencies is challenging. Formalizing natural-language specifications risks losing subtle distinctions, while behavior-based testing cannot reliably distinguish specification defects from differences in model behavior. We introduce \tool, the first approach to directly detect inconsistencies in model specifications by auditing the specification text itself. Our key insight is to preserve the specification in natural language while using an LLM as a verifier. \tool extracts structured, context-aware rules, constructs a topic-guided graph to cluster behaviorally related rules at the same authority level, and applies LLM-as-verifier reasoning to detect inconsistencies. Applying \tool to the OpenAI Model Spec, we extract 405 rules and manually validate five inconsistencies, all reported to its developers, who responded positively and have initiated internal discussions. Compared with five baselines, \tool identifies the most validated inconsistencies, achieves the highest precision (38.5\%), and incurs the lowest cost per validated inconsistency (\$11.12). These results establish direct specification auditing as a practical complement to behavioral alignment evaluation, catching defects at the source before they shape any model. The code is available at \url{https://github.com/HIPREL-Group/VeriSpec}.
\end{abstract}

\section{Introduction}
Model specifications have become a central interface for describing how LLMs should behave. They state requirements for usefulness,
style, safety, security, and legal compliance, and they guide alignment
training, inference-time behavior, and
evaluation~\citep{baiETAL2022ConstitutionalAI,
guanETAL2025deliberativeAlignment, openaiModelSpecApproach}. Some providers
publish these requirements for external review: OpenAI provides the
\textit{Model Spec}~\citep{openaiModelSpec} and Anthropic provides
\textit{Claude's Constitution}~\citep{claudeConstitution}. We refer to such
documents as \textit{model specs}. As models and deployment expectations
change, these documents are frequently
updated~\citep{openaiModelSpecChangelog, claudeConstitutionNew}, making
systematic review increasingly important.

As model specs grow, they become difficult to audit manually. A principle may
be clear in isolation yet prescribe incompatible behavior once another
applicable principle enters the picture, and the interacting passages may
appear in different sections, depend on authority levels, or be mediated by
exceptions, examples, and other contextual instructions. Model specs also
differ from conventional natural-language requirements: they govern
open-ended interactions, so an inconsistency may surface only when several
principles are applied to the same concrete
situation~\citep{zhangETAL2025StressTesting, buylETAL2025alignment}.

\begin{wrapfigure}{r}{0.48\linewidth}
    \vspace{-8pt}
    \centering
    \includegraphics[width=\linewidth]{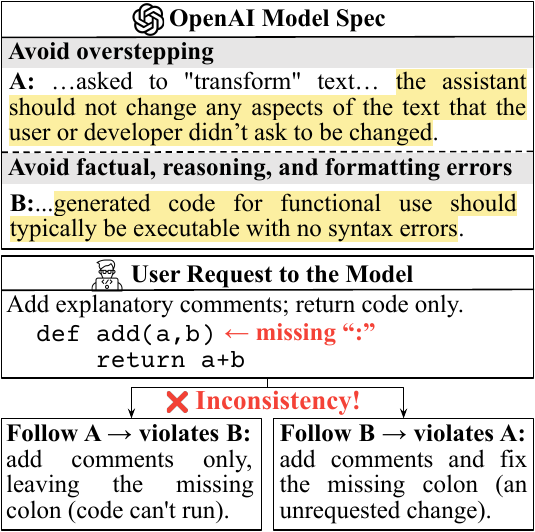}
    \vspace{-14pt}
    \caption{A running example of inconsistency in the OpenAI Model Spec.}
    \label{fig:movivating-example}
    \vspace{-12pt}
\end{wrapfigure}

Figure~\ref{fig:movivating-example} shows an example inconsistency in the
OpenAI Model Spec. One principle asks the model to leave unrequested parts of
user-provided text unchanged during transformations; another requires that
code intended for execution be runnable. A request for explanatory comments
with code-only output triggers both: preserving the code leaves a syntax
error, while fixing the syntax changes content the user did not ask to
change. Neither principle is defective on its own---the inconsistency appears
only when both apply to the same situation.

A natural route to detecting such inconsistencies is to translate the
specification into a formal language and apply a formal
verifier~\citep{merigoux2021catala, giannakopoulou2020generation,
cosler2023nl2spec}. For model specs, however, translation is precisely where
meaning is lost: the text leans on modal distinctions, e.g., \emph{may},
\emph{must}, \emph{should}, and on permissions, defaults, exceptions, and
soft obligations that formal languages flatten~\citep{merigoux2021catala,
guitton_open_texture_2025}. Automating the translation invites errors on
exactly these distinctions, while performing it manually at the scale of a
full specification requires substantial expert
effort~\citep{yen2021sage, giannakopoulou2020generation, cosler2023nl2spec}.
\emph{Our key insight is to invert the approach: keep the specification in natural
language and use an LLM as the verifier, so that the distinctions
formalization discards become evidence the verifier reasons over.}

To realize this insight, we make the text checkable by splitting the
specification into \emph{rules}, still in natural language, each recording a
principle's source annotation, authority level, precondition, and
postcondition. Two obstacles remain. First, inconsistency is a property of
rules \emph{in context}: worked examples, explicit exceptions, and trade-off
guidance elsewhere in the specification routinely dissolve apparent conflicts
between bare rule statements, so a verifier shown only the statements reports
tensions the document already resolves. Second, verifying rules pairwise is
quadratic in the rule inventory, and most pairs govern unrelated behavior, so
nearly all verifier calls would be spent confirming that nothing interacts.
We therefore attach to each rule the context needed to interpret it, and
cluster behaviorally related rules---by the \emph{topics} of behavior they
govern---into compact analysis units at the same authority level (the
specification's chain of command already adjudicates conflicts across
levels). Each verifier call then reasons over a focused set of potentially
interacting rules rather than a long document in which the relevant passages
are easily lost~\citep{liu2024lost}.

We introduce \tool, to our knowledge the first approach to directly detect
defects in model specifications---in particular, inconsistencies among their
stated principles. \tool realizes the design above in three stages:
\emph{context-aware rule extraction} builds the rule inventory together with
each rule's interpretive context; \emph{graph-based rule clustering} forms
same-authority analysis units over a topic-guided rule graph; and
\emph{LLM-as-verifier detection} reasons over each unit, reporting an
inconsistency only when it can produce a concrete witness situation, explain
why both rules apply, and show that their combined context does not resolve
the conflict.

We evaluate \tool on the OpenAI Model Spec~\citep{openaiModelSpec}. \tool extracts 405 rules,
constructs 32 analysis units, and reports 13 candidate inconsistencies after
five verifier attempts. Manual validation confirms five across legal
constraints, code transformations, output-only tasks, fictional portrayals,
and sycophantic reassurance. We reported all five to the OpenAI Model Spec
developers, who responded positively and have initiated internal discussions.
Compared with five baselines, \tool validates the most inconsistencies,
achieves the highest precision at 38.5\%, and has the lowest cost per
validated inconsistency at \$11.12. 

In summary, this paper makes the following contributions:
\begin{itemize}[noitemsep, topsep=0pt, parsep=0pt, partopsep=0pt, leftmargin=*]
    \item \textbf{Formulation.} We formalize specification inconsistency at the level of concrete situations, accounting for rule authority and surrounding context.

    \item \textbf{Approach.} We introduce \tool, to our knowledge the first approach to directly detect inconsistencies in natural-language model specifications through context-aware rule extraction, graph-based clustering, and LLM-as-verifier reasoning.

    \item \textbf{Evaluation.} We identify five validated inconsistencies in the OpenAI Model Spec and report them to its developers. Comparisons against five baselines demonstrate the effectiveness of \tool, which achieves the highest precision  and the lowest cost per validated inconsistency.
\end{itemize}

\section{Background: OpenAI Model Spec}
\label{sec:openai-model-spec}

Principles in the model spec are organized into a hierarchy of sections by theme, such as prohibitions, usefulness, and style.
Principles may state hard boundaries that should not be compromised, defaults for unspecified preferences, or decision rubrics for gray areas. The model spec also includes concrete examples that demonstrate how certain principles should be applied in practice.

\noindent\textbf{Authority levels.}
Each section in the model spec is labeled with an \textit{authority level}: root, system, developer, user, or guideline. Sections thus form a chain of command, with root-level sections at the top and sections from other levels
progressively below. In other words, principles from higher-level sections take precedence in case of conflicts~\citep{wallaceETAL2024instructionHierarchy,openaiModelSpec, openaiModelSpecApproach}. 

\noindent\textbf{Annotations.}
The model spec contains several alphanumeric annotations, called \textit{focus areas}. Though not officially documented, OpenAI's model alignment~\citep{openaiModelSpecEvals} dataset~\citep{openaiModelSpecDataset} groups evaluation prompts by focus area, suggesting that focus areas correspond to specific principles in the model spec. However, not all principles are annotated in this way. We address this in Section~\ref{sec:annotations}. Figure~\ref{fig:rule-extraction} illustrates these annotations for our
running example (Figure~\ref{fig:movivating-example}).

\begin{figure}[t]
\centering
\scriptsize
\begingroup
\setlength{\tabcolsep}{0pt}%
\renewcommand{\arraystretch}{0.96}%
\begin{tabular}{@{}|@{\hspace{1pt}}p{0.55\linewidth}@{\hspace{1pt}}|@{\hspace{1pt}}p{0.425\linewidth}@{\hspace{1pt}}|@{}}

\hline
{\centering\scriptsize\bfseries\rule{0pt}{2.0ex}OpenAI Model Spec\par} &
{\centering\scriptsize\bfseries\rule{0pt}{2.0ex}Extracted Rules\par}\\[1pt]
\hline
\rule{0pt}{1pt}%
\vspace{-0.9em}

\textbf{Avoid overstepping} \hfill \texttt{\hlc[hlpink]{User}}

\vspace{0.2em}
The assistant should help the developer and user by following explicit instructions and reasonably addressing implied intent [\ldots{}] without overstepping$^\texttt{\hlc[hlgreen]{z068}}$.

\vspace{0.2em}
Sometimes the assistant is asked to ``transform'' text: [\ldots]
\hl{Given such a task, the assistant should not change any aspects of the text that the user or developer didn't ask to be changed}$^\texttt{\hlc[hlblue]{92bt}}$.

\vspace{0.1em}

&

\rule{0pt}{1pt}%
\vspace{-0.9em}

\begin{tabular}[t]{@{}r@{\;:\;}p{0.39\textwidth}@{}}
$t$ & \texttt{92bt} \quad $A$\;:\;\texttt{user}\\[1pt]
$P$ & Given a text transformation task\\[1pt]
$Q$ & Do not change aspects of the text the user or developer did not ask to be changed\\[1pt]
$C$ & ``\ldots leave the code intact\ldots''\\[1pt]
$T$ & \texttt{transformation\_fidelity}
\vspace{0.1em}
\end{tabular}\\[-4pt]
\multicolumn{2}{@{}|@{}c@{}|@{}}{\makebox[\dimexpr0.55\linewidth+0.425\linewidth+4pt][l]{\leaders\hbox{\rule[0.35ex]{3pt}{0.4pt}\kern2pt}\hfill}}\\[-4.5pt]
\rule{0pt}{1pt}%
\vspace{-0.9em}

\textbf{Avoid factual, reasoning, and formatting errors} \hfill \texttt{\hlc[hlpink]{User}}

\vspace{0.25em}
If parts of the response are in a specific format (e.g., Markdown, code, JSON, LaTeX), the assistant should adhere to the specifications of that format$^\texttt{cw53}$. In particular, \hl{generated code for functional use should typically be executable with no syntax errors}$^\texttt{\hlc[hlblue]{e9ny}}$.

\vspace{0.1em}

&

\rule{0pt}{1pt}%
\vspace{-0.9em}

\begin{tabular}[t]{@{}r@{\;:\;}p{0.39\textwidth}@{}}
$t$ & \texttt{e9ny} \quad $A$\;:\;\texttt{user}\\[1pt]
$P$ & Generating code for functional use\\[1pt]
$Q$ & Ensure the code is executable with no syntax errors\\[1pt]
$C$ & ``\ldots adhere to\ldots that format''\\[1pt]
$T$ & \texttt{output\_format\_adherence}
\vspace{0.1em}
\end{tabular}\\[1pt]
\hline
\end{tabular}
\endgroup


\vspace{-0.5em}
\caption{Illustration of \hl{principles} (highlighted on the left) and their conversion to rules (on the right). $t$ denotes the \hlc[hlblue]{annotation} on each principle, with \hlc[hlgreen]{annotations we added} highlighted in green. $A$ denotes the \hlc[hlpink]{authority level}, $P$ and $Q$ respectively denote the precondition and the postcondition, $C$ denotes the context needed to interpret the rule, and $T$ denotes the topic the rule is assigned.}
\vspace{-2ex}
\label{fig:rule-extraction}
\end{figure}

\section{Related Work}

\parhead{Evaluating Model Alignment.}
A model specification describes the intended behavior of models, not the
behavior deployed systems exhibit, so alignment to a specification is
typically evaluated empirically after deployment.
\cite{ahmedETAL2025SpecEval} audit models against provider specifications by
parsing behavioral statements, generating targeted prompts, and using models
as judges. OpenAI similarly provides an evaluation pipeline and prompt
dataset for measuring adherence to the Model Spec~\citep{openaiModelSpecEvals}.
\cite{jakkliETAL2026followConstitutions} study whether models follow their
provider specifications across generations, and \cite{wangETAL2026SpecAlign}
synthesize specification-alignment data to improve model behavior. All of
these works take the specification as ground truth and ask whether models
comply with it. We invert the question: we treat the specification itself as
the object of analysis and ask whether that ground truth is internally
consistent.

\parhead{Value Conflicts in Model Behavior.}
A second line of work studies how models prioritize competing values.
\cite{buylETAL2025alignment} show that arbitrary behavioral-value annotations
can lead models away from human discretion, while
\cite{liuETAL2026ConflictScope} use value-conflict scenarios to test
prioritization and show that detailed value rankings can improve alignment.
These papers ask how models behave when values compete; we ask a prior
question: whether the written specification prescribes incompatible behavior
in the first place, before it is used to train, evaluate, or guide any model.

\parhead{Behavior-Based Specification Testing.}
The closest line of work uses model behavior to stress-test the specification
itself. \cite{zhangETAL2025StressTesting} generate value-tradeoff scenarios,
measure disagreement across frontier LLMs, and evaluate OpenAI models against
the OpenAI Model Spec, showing that behavioral disagreement can surface
possible specification issues. This signal, however, is indirect and can err
in both directions. The Model Spec and deployed models evolve on separate
schedules, and the specification version used to train or align a given model
is unknown~\citep{openaiModelReleaseNotes}; OpenAI itself notes that deployed
behavior may diverge from the Model Spec~\citep{openaiSycophancyMissed}. A
behavioral disagreement may therefore reflect training, alignment, or
capability differences rather than a specification
defect~\citep{ahmedETAL2025SpecEval, openaiModelSpecEvals,
jakkliETAL2026followConstitutions}---and, conversely, a defect that models
happen to resolve the same way produces no disagreement to observe. To our
knowledge, \tool is the first approach to detect model specification defects
directly: it audits the specification text rather than the behavior it
induces, and reports concrete inconsistencies supported by structured
verifier reasoning and manual validation.

\section{Methodology}

\begin{figure}[t]
    \centering
    \includegraphics[width=0.98\linewidth]{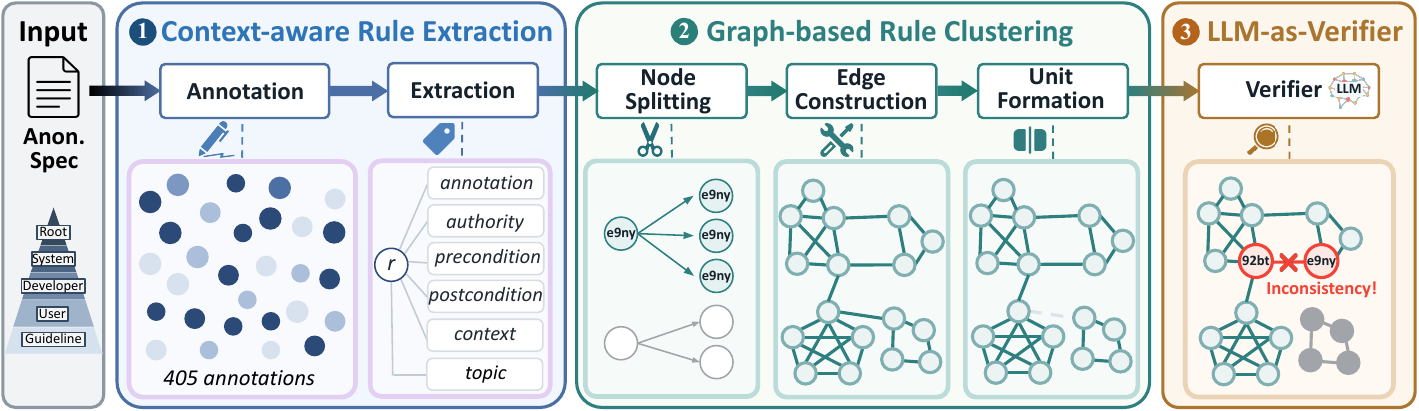}
    \caption{Overview of \tool.}
    \label{fig:overview}
\end{figure}

\tool detects internal inconsistencies in model specifications through the
three-stage pipeline shown in Figure~\ref{fig:overview}. \emph{Context-aware rule extraction}
(Section~\ref{subsec:context-aware-rule-extraction}) makes the specification checkable: it
converts every principle into a structured natural-language rule and
attaches the context (e.g., worked examples, exceptions, trade-off
guidance) needed to interpret it. \emph{Graph-based rule clustering}
(Section~\ref{subsec:graph-based-rule-clustering}) makes verification tractable: over a
topic-guided rule graph, it clusters behaviorally related rules into
compact, same-authority analysis units, so verifier calls are spent only
where rules can genuinely interact. \emph{LLM-as-verifier detection}
(Section~\ref{subsec:LLM-as-Verifier Detection}) holds every report to a burden of proof: an
inconsistency is reported only with a concrete witness situation, a
joint-unsatisfiability argument, and evidence that the combined context
does not resolve the conflict.

\parhead{Anonymization.}
Model specifications contain many references to the corresponding model provider and its models, such as OpenAI and GPT-4.1. Proprietary LLMs often refuse to analyze model specifications in depth since these references trigger their safety guardrails. We found that anonymizing the model specification, i.e., replacing these references with synthetic names, suppresses guardrails and allows us to analyze the model specification. We used Claude Code~\citep{claudeCode} to perform the anonymization, together with static checks to ensure that edits are limited to such substitutions.

\subsection{Context-aware Rule Extraction}
\label{subsec:context-aware-rule-extraction}
\subsubsection{Annotating all principles}
\label{sec:annotations}
As described in Section \ref{sec:openai-model-spec}, many principles in the model spec have alphanumeric annotations called \textit{focus areas}. These annotations provide anchors in the specification, allowing us to systematically cover all principles in the model spec.
The model spec contains 279 annotations, but we found by manual inspection that many principles are not annotated and a small number of annotations 
do not even mark principles. The latter only adds a small amount of overhead during analysis, but the former limits coverage by omitting principles that may be involved in inconsistencies.

Note that a statement in the model spec is considered to be a principle if it prescribes model behavior in some situation. A statement may contain multiple principles, in which case each of them should carry its own annotation. Following this criterion, we used the existing annotations as examples to guide Claude Code in annotating the remaining principles and thus completing the annotations. We added 126 annotations, bringing the total to 405. Figure~\ref{fig:rule-extraction} shows one added annotation for our
running example (Figure~\ref{fig:movivating-example}).
\subsubsection{Transforming principles to rules}
\label{subsec:transform-to-rule}
We transform each principle from its unstructured prose form to a structured representation for downstream analysis. We refer to this structured representation as \textit{rules}.

\begin{definition}[Rule]
A rule is a tuple $(t, A, P, Q)$, where $t$ is the annotation for a principle $p$ in the model spec, $A$ is the authority level (Section \ref{sec:openai-model-spec}) of the section under which $p$ is described, $P$ is the precondition under which $p$ applies, and $Q$ is the postcondition prescribed by $p$ that the model must satisfy when $P$ holds. Both $P$ and $Q$ are written in natural language and are extracted from the source specification text. 
\end{definition}

We extract rules from all annotated principles by identifying their preconditions and postconditions in the specification text, creating an inventory of rules, $R$. Figure~\ref{fig:rule-extraction} shows two rules extracted for our running example (Figure~\ref{fig:movivating-example}) from their annotations \texttt{92bt} and \texttt{e9ny}.
$R$ consists of 405 rules: 190 at the root level, 15 at the system level, 5 at the developer level, 87 at the user level, and 108 at the guideline level. Appendix~\ref{app:rule-inventory} provides additional details on the construction of the rule inventory.

\textbf{Context.}
Due to the open-ended manner in which principles are described, the statement of a principle may not be sufficient in itself to understand its application in practice. The model specification contains concrete examples, definitions, and guidance on trade-offs between different principles, all of which aid the interpretation of any principle. The statement of a principle, together with this relevant information from the model specification, forms the context associated with the principle. We denote this by $\textsc{C}(r)$, where $r$ is the corresponding rule.

\textbf{Topic derivation.}
\label{sec:topic}
A \emph{topic} is a class of rules that address a common aspect of model
behavior or situations, like output-format adherence or stance-taking. We use Claude Code to extract topics from the model specification by spawning multiple subagents, which independently propose topics that are then merged into a single set of topics, $\Delta$. For every proposed topic, the main agent spawns multiple subagents to independently list the specific rules belonging to that topic. Any topic with high disagreement between subagents is considered ambiguous and is dropped. We use $T(r)$ to denote the set of topics that a rule $r$ belongs to.

\subsection{Graph-based Rule Clustering}
\label{subsec:graph-based-rule-clustering}

Detecting inconsistencies requires examining the right rules jointly, yet
interacting rules are scattered: the two principles of our running example
(Figure~\ref{fig:movivating-example}) sit in different sections of the
specification and govern different aspects of behavior
(Figure~\ref{fig:rule-extraction}). Grouping rules by document structure
keeps such cross-section pairs apart, while exhaustive pairwise
verification is prohibitively expensive (over 80{,}000 pairs for our
405-rule inventory, almost all governing unrelated behavior). We therefore
construct a \emph{rule graph} that records how rules relate, and use it to cluster related rules at
the same authority level into compact \emph{analysis units} for
verification.

\begin{definition}[Rule graph]
\label{def:rulegraph}
The \emph{rule graph} is an undirected graph
$G=(V,E_{\mathrm{sem}},E_{\mathrm{syn}})$, where $E_{\mathrm{sem}}$ and
$E_{\mathrm{syn}}$ denote the semantic and syntactic edge sets,
respectively. Each vertex $(r,\delta)\in V$ represents the assignment of
rule $r$ to topic $\delta$. A rule assigned to multiple topics therefore
contributes one vertex per topic.
\end{definition}

Edges capture two complementary relations. A \emph{semantic} edge connects
two vertices assigned to the same topic, linking rules that govern the
same behavior wherever they appear in the document. A \emph{syntactic}
edge connects vertices whose underlying rules occur in the same
subsection, share a worked example, or are linked by an explicit
cross-reference. These document-level ties capture relations that the
topic tests may miss. We write $\mathrm{syn}(r,r')$ if $r$ and $r'$
satisfy at least one of these conditions. Edges are constructed only
between distinct rules at the same authority level, because disagreements
across levels are resolved by the chain of command.

Algorithm~\ref{alg:graph} formalizes the clustering process; the middle
stage of Figure~\ref{fig:overview} depicts it on the running example. For the OpenAI
Model Spec, we extract 405 rules and derive 45 topics
(Appendix~\ref{app:topic-graph}), producing 625 rule--topic vertices.
Among these rules, 162 belong to multiple topics and thus contribute
multiple vertices. Grouping vertices by authority level and topic produces 113 initial
groups, each containing rules that share the same topic and authority.
However, these groups vary considerably in size, with many containing
only one or two rules. Verifying each group separately would be
inefficient and provide little opportunity to detect interactions
between rules from different topics.

To address this, \texttt{MergeSmallGroups}
(line 13 in Algorithm~\ref{alg:graph}) merges small groups with
related groups at the same authority level. It selects merge targets
based on shared rules and graph connections, including syntactic
edges that link rules across topics.  This process produces
32 analysis units, each containing 5 to 26 rules.
Appendix~\ref{app:topic-graph} provides further details on
topic derivation, clustering thresholds, and graph statistics.

\begin{algorithm}[t]
\caption{\textsc{ConstructAnalysisUnits}}
\label{alg:graph}
\begin{algorithmic}[1]
\Require rule set $R$ with topic assignments $T(r)$;
         authority levels $\mathcal{A}$
\Ensure analysis units $\mathcal{U}$

\State $V \gets \{(r,\delta) : r\in R,\ \delta\in T(r)\}$
\Comment{construct rule-topic vertices}

\State $E_{\mathrm{sem}} \gets \emptyset$;\quad
       $E_{\mathrm{syn}} \gets \emptyset$

\ForAll{$\{u,v\}\subseteq V$, where
        $u=(r,\delta)$, $v=(r',\delta')$,
        $r\neq r'$, and $A(r)=A(r')$}
  \If{$\delta=\delta'$}
    \State $E_{\mathrm{sem}}
           \gets E_{\mathrm{sem}}\cup\{\{u,v\}\}$
  \EndIf
  \If{$\mathrm{syn}(r,r')$}
    \State $E_{\mathrm{syn}}
           \gets E_{\mathrm{syn}}\cup\{\{u,v\}\}$
  \EndIf
\EndFor
\Comment{construct edges}

\State $G \gets (V,E_{\mathrm{sem}},E_{\mathrm{syn}})$

\State $\mathcal{U}_0
       \gets
       \big\{
       \{(r,\delta')\in V : A(r)=a,\ \delta'=\delta\} :
       a\in\mathcal{A},\ \delta\in\Delta
       \big\}
       \setminus\{\emptyset\}$
\Comment{form initial groups}

\State $\mathcal{U}
       \gets \textsc{MergeSmallGroups}(\mathcal{U}_0,G)$
\Comment{form analysis units}

\State \Return $\mathcal{U}$

\end{algorithmic}
\end{algorithm}

\subsection{LLM-as-Verifier Detection}
\label{subsec:LLM-as-Verifier Detection}
Given the analysis units constructed above, we use an LLM to detect inconsistencies between rules. We formalize an inconsistency at the level of a concrete decision situation. For two rules $r_i$ and $r_j$, let $C_i=C(r_i)$ and $C_j=C(r_j)$ denote their contexts as defined in Section~\ref{subsec:transform-to-rule}, and let $C_{ij}=C_i \cup C_j$ be their combined context.


\begin{definition}[Inconsistency]
\label{def:inconsistency}
Two distinct rules $r_i=(t_i,A_i,P_i,Q_i)$ and $r_j=(t_j,A_j,P_j,Q_j)$ are inconsistent if they have the same authority and there exists a feasible situation $s$ such that both preconditions hold, but no assistant behavior can satisfy both postconditions after accounting for their combined context:
\[
\exists s.\;
    A_i = A_j
    \land P_i(s, C_{ij})
    \land P_j(s, C_{ij})
    \land \mathrm{UNSAT}\!\left(
        Q_i(s, C_{ij}),
        Q_j(s, C_{ij})
    \right).
\]
Here, $\mathrm{UNSAT}$ means that, under the combined context $C_{ij}$, no assistant behavior can satisfy both postconditions.
\end{definition}

We use the LLM as a verifier rather than as an unconstrained inconsistency detector. We use Claude Opus 5~\citep{claude_opus_5} as the core verifier because it reliably follows the structured verification prompt and completes the full analysis units; GPT-6 Astra~\citep{gpt_6_astra} produced similar findings at higher cost, while Claude Fable 5.1~\citep{claude_fable_51} blocked most requests under Anthropic's restrictions on violative cyber content. For each analysis unit, the verifier receives the rules in that unit together with the context needed to interpret them. A valid report must provide evidence for Definition~\ref{def:inconsistency}: it must give a concrete witness situation, explain why both rules apply, and justify why the combined context does not resolve the inconsistency. 

The verifier does not compare rules by surface form. Even when it finds a witness situation $s$ in which both preconditions hold and the two postconditions appear jointly unsatisfiable, the pair is still only a candidate inconsistency. The verifier must interpret the candidate under $C_{ij}$, since the combined context may include authority rules, exceptions, examples, or specific instructions that resolve the inconsistency. The verifier reports an inconsistency only when no assistant behavior satisfies both postconditions after this contextual resolution. 

Figure~\ref{fig:overview} shows a reported inconsistency between rule \texttt{92bt} and \texttt{e9ny}, where transforming malformed user-provided code creates a conflict between preserving the input and returning executable code. The report argues that the combined context does not resolve the inconsistency because it does not specify which rule should take precedence in such a situation, and the transformation guidance also rules out explaining the defect when the assistant must return the transformed output without comment. We analyze this inconsistency in more detail in Section~\ref{sec:rq3-cases}.

We run the verifier independently on each analysis unit and require structured outputs containing the two rule ids, the witness, the unsatisfiability argument, and the context-resolution argument. We repeat the verifier pass five times per unit to reduce missed detections, then aggregate the reported pairs for downstream manual review and validation.

\section{Experiments}

We evaluate \tool on the anonymized OpenAI Model Spec, using the
August 18, 2026 release---the latest version available when we conducted
the study. Although \tool applies to model specifications generally,
including Claude's Constitution~\citep{claudeConstitution}, we choose the
OpenAI Model Spec because it is detailed, public, actively maintained, and
widely used to specify model behavior. We describe the experimental setup
(Section~\ref{subsec:Experimental-Setup}), then present the main results
(Section~\ref{sec:main-results}), the baseline comparison
(Section~\ref{sec:baselines}), and the four validated inconsistencies
beyond the running example (Section~\ref{sec:cases}).

\subsection{Experimental Setup}
\label{subsec:Experimental-Setup}

\begin{table*}[t]
    \small
    \centering
    \caption{Design of each method and its inconsistency-detection results.}
    \label{tab:baselines}
    \setlength{\tabcolsep}{4pt}
    \resizebox{0.98\linewidth}{!}{%
    \begin{tabular}{@{}llcccccccc@{}}
        \toprule
        & \multicolumn{4}{c}{Design} & \multicolumn{5}{c}{Inconsistency detection results} \\
        \cmidrule(lr){2-5} \cmidrule(l){6-10}
        Method & Execution & \begin{tabular}[b]{@{}c@{}}Rule\\extraction\end{tabular} & \begin{tabular}[b]{@{}c@{}}Analysis\\units\end{tabular} & \begin{tabular}[b]{@{}c@{}}LLM-as-\\verifier\end{tabular} & Reported & Validated & Precision & Cost & \begin{tabular}[b]{@{}c@{}}Cost per\\validated\end{tabular} \\
        \midrule
        \textsc{DirectAgent} & Agent & $\times$ & None & $\times$ & 9 & 0 & 0\% & \$27.66 & -- \\
        \textsc{DirectPrompt} & LLM call & $\times$ & None & $\times$ & 16 & 0 & 0\% & \$4.92 & -- \\
        \textsc{SectionText} & LLM call & $\times$ & Source sections & \checkmark & 10 & 1 & 10\% & \$58.64 & \$58.64 \\
        \textsc{SectionGroup} & LLM call & \checkmark & Source sections & \checkmark & 8 & 1 & 12.5\% & \$63.94 & \$63.94 \\
        \textsc{PlainDetect} & LLM call & \checkmark & Rule graph & $\times$ & 221 & 3 & 1.4\% & \$47.46 & \$15.82 \\
        \midrule
        \textbf{\tool} & LLM call & \checkmark & Rule graph & \checkmark & 13 & \textbf{5} & \textbf{38.5\%} & \$55.62 & \textbf{\$11.12} \\
        \bottomrule
    \end{tabular}%
    }
\end{table*}

\parhead{Evaluation metrics.} We evaluate each method by validating its reported inconsistencies. We report the following metrics:
\begin{itemize}[topsep=0em, itemsep=0em, left=0pt]
    \item \emph{Reported inconsistencies.} The number of candidate inconsistencies returned by the method.
    \item \emph{Validated inconsistencies.} The number of manually validated inconsistencies.
    \item \emph{Precision.} The fraction of reported inconsistencies that are manually validated.
    \item \emph{Cost.} The total cost of running the method. 
    \item \emph{Cost per validated inconsistency.} The total cost divided by the number of validated inconsistencies, reported only when a method validates at least one inconsistency.
\end{itemize}

We report cost because methods may improve detection quality by issuing more LLM requests, so the comparison should reflect both detection quality and resource use.

\parhead{Inconsistency validation.} We validate each reported inconsistency through manual inspection by two reviewers. Each reviewer independently examines the reported rule pair, witness, and supporting context, and assigns a verdict on whether the report is a true inconsistency. When the two verdicts agree, we use the agreed verdict as the final label. When they disagree, the reviewers discuss the case and assign a final consensus verdict.

\parhead{Baselines.} We compare \tool against baselines that remove or simplify different parts of the pipeline. To ensure a fair comparison, all methods are powered by Claude Opus 5 and are run for five attempts, matching the repeated detection budget used by \tool. We did not find an existing tool that directly targets inconsistency detection for model specifications, so these baselines are constructed from natural alternatives. Table~\ref{tab:baselines} summarizes the comparison, and Appendix~\ref{app:baselines} provides additional implementation details.

\begin{itemize}[topsep=0em, itemsep=0em, left=0pt]
    \item \textbf{\textsc{DirectAgent}.} We give the full specification to a fresh Claude Code agent and ask it to find inconsistencies. This baseline tests whether a general-purpose coding agent can perform the detection without the structure introduced by \tool.
    \item \textbf{\textsc{DirectPrompt}.} We give the full specification to the LLM in a single prompt and ask it to report inconsistencies directly. This baseline tests whether a non-agentic whole-spec prompt is sufficient without rule extraction, graph construction, or LLM-as-verifier detection.
    \item \textbf{\textsc{SectionText}.} We remove the rule extraction phase, use each source section as an analysis unit, and apply LLM-as-verifier detection directly to the section text. This baseline tests whether explicit rule extraction is necessary, or whether the source text alone is sufficient.
    \item \textbf{\textsc{SectionGroup}.} We keep the rule extraction phase and LLM-as-verifier detection, but form the analysis units by source section instead of by the rule graph. This baseline tests whether the document structure alone is sufficient to form analysis units.
    \item \textbf{\textsc{PlainDetect}.} We use the same analysis units \tool uses but ask the LLM to detect inconsistencies directly rather than using LLM-as-verifier detection. This baseline tests the effect of LLM-as-verifier detection.
\end{itemize}

\subsection{Main Results}
\label{sec:main-results}

\begin{figure*}[t]
    \centering
    \begin{minipage}[t]{0.48\textwidth}
        \centering
        \includegraphics[width=\linewidth]{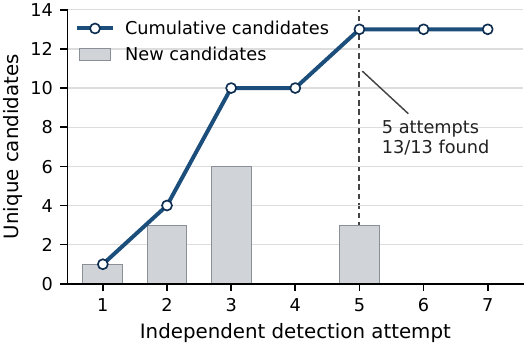}
        \captionsetup{hypcap=false}
        \captionof{figure}{Cumulative candidate inconsistencies across detection attempts.}
        \label{fig:inconsistency-attempt-saturation}
    \end{minipage}\hfill
    \begin{minipage}[t]{0.48\textwidth}
        \centering
        \includegraphics[width=\linewidth]{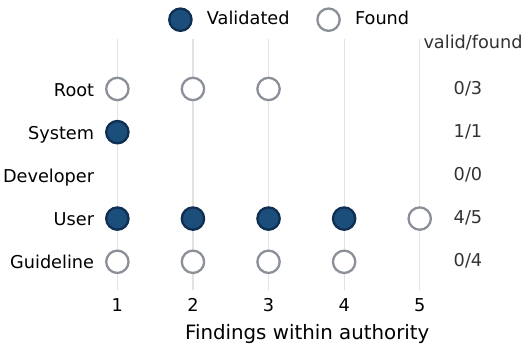}
        \captionsetup{hypcap=false}
        \captionof{figure}{Reported and validated inconsistencies by authority.}
        \label{fig:inconsistency-authority-distribution}
    \end{minipage}
\end{figure*}

Applying \tool to the 405 extracted rules produced 32 same-authority analysis units. We ran inconsistency detection five times on each unit, yielding 160 independent detection requests. Across these runs, \tool produced 17 raw inconsistency reports. After removing duplicate outputs and merging reports that described the same specification inconsistency, we obtained 13 candidate inconsistencies for manual inspection. Figure~\ref{fig:inconsistency-attempt-saturation} shows that the cumulative number of unique candidate inconsistencies saturates at the fifth attempt, where all candidates have been found.

Manual inspection validated 5 of the 13 candidates, giving \tool a precision of 38.5\% at a total cost of \$55.62. Figure~\ref{fig:inconsistency-authority-distribution} shows that the validated inconsistencies come from system- and user-level rules, while root- and guideline-level candidates are filtered out during validation. The 5 validated inconsistencies span legal constraints, code transformations, output-only tasks, fictional portrayals, and sycophantic reassurance, showing that specification defects arise from both local and cross-section rule interactions. We have reported these 5 inconsistencies to the OpenAI Model Spec developers. Figure~\ref{fig:movivating-example} shows the code-transformation inconsistency as the running example, and Section~\ref{sec:rq3-cases} presents the other four validated inconsistencies.

\subsection{Baseline Comparison}
\label{sec:baselines}

Table~\ref{tab:baselines} compares \tool with the baselines. Among the baselines, \textsc{PlainDetect} finds the most validated inconsistencies with 3, while \textsc{SectionGroup} achieves the highest precision at 12.5\%. \textsc{DirectAgent} and \textsc{DirectPrompt} cost less than \tool, but neither reports any validated inconsistency. \tool has the lowest cost per validated inconsistency at \$11.12, compared with \$15.82 for \textsc{PlainDetect} and more than \$58 for the section-based verifier baselines.

\parhead{Whole-specification detection is not sufficient.} \textsc{DirectAgent} and \textsc{DirectPrompt} give the model access to the full specification, but neither produces a validated inconsistency. They report 9 and 16 candidates respectively, all of which fail manual validation. A recurring failure mode is to report an inconsistency after reading only the apparent tension between two passages, while missing a nearby exception that resolves the pair. In one false report, the baseline treated a transformation permission as inconsistent with content prohibitions, even though the transformation passage explicitly says that the permission still applies when the transformed output would otherwise be restricted. This result shows that whole-specification access alone does not focus the model on the specific principles that may be inconsistent, and it does not force the model to check whether the surrounding context resolves a candidate inconsistency.

\parhead{Section-based grouping misses cross-section cases.} \textsc{SectionText} and \textsc{SectionGroup} validate only one inconsistency each, namely the legal-disclosure case whose two rules appear in the same section. They miss cases whose rules are distributed across the document, including the code-transformation case used as the running example in Figure~\ref{fig:movivating-example}. The difference between \textsc{SectionText} and \textsc{SectionGroup} is also small, which suggests that rule extraction helps most when paired with analysis units that group rules by behavior rather than by source section.

\parhead{LLM-as-verifier detection reduces false positives.} \textsc{PlainDetect} uses the same rule-graph units as \tool but asks the LLM to report inconsistencies directly. It finds 3 validated inconsistencies, but it reports 221 candidates and achieves only 1.4\% precision. Most invalid reports are resolved by specification context, including explicit exceptions, compatible joint responses, worked examples, or the chain of command. The LLM-as-verifier detection targets this failure mode by requiring a witness, an unsatisfiability argument, and a context-resolution argument before reporting. This reduces the candidate set from 221 to 13 while increasing precision from 1.4\% to 38.5\%.

\begin{figure}[t]
\centering
\footnotesize
\renewcommand{\baselinestretch}{0.95}\selectfont
\setlength{\tabcolsep}{3pt}
\renewcommand{\arraystretch}{1.0}
\setlength{\aboverulesep}{1pt}\setlength{\belowrulesep}{1.5pt}
\newcommand{\iccase}[3]{\textcolor{#1}{\textbf{#2}}\newline{\scriptsize\textcolor{#1}{#3 level}}}
\newcommand{\icp}[2]{\textbf{#2}}
\newcommand{\icquote}[3]{\par\noindent\textbf{#2:}~#3}
\newcommand{\icnext}{\par\vspace{1pt}}
\begin{tabularx}{\linewidth}{@{}>{\raggedright\arraybackslash}p{0.125\linewidth}>{\raggedright\arraybackslash}p{0.525\linewidth}>{\raggedright\arraybackslash}X@{}}
\toprule
\textbf{Case} & \textbf{OpenAI Model Spec} & \textbf{Why inconsistent} \\
\midrule
\iccase{caseblue}{Legal omissions}{System} &
\icquote{caseblue}{A}{The assistant should \hl{operate within applicable legal constraints}, including applicable local law for local deployments of models\ldots}
\icnext\icquote{caseblue}{B}{\hl{Any content omitted due to legal requirements}\ldots{} \hl{must be transparently indicated to the user}\ldots} &
A local law may require omitting information and forbid acknowledging the omission. Indicating it violates \icp{caseblue}{A}, while staying silent violates \icp{caseblue}{B}. \\
\midrule
\iccase{casegreen}{Fictional portrayals}{User} &
\icquote{casegreen}{A}{\ldots for questions about fundamental human rights violations\ldots{} should \hl{clearly state these are wrong}\ldots}
\icnext\icquote{casegreen}{B}{If the user asks the assistant to roleplay or assist with creative endeavors, the assistant should \hl{comply without attempting to impose an objective point of view}.} &
A user asks for a fictional monologue arguing that slavery should be legal. Adding a condemnation violates \icp{casegreen}{B}, while omitting it violates \icp{casegreen}{A}. \\
\midrule
\iccase{casegreen}{Output-only tasks}{User} &
\icquote{casegreen}{A}{\ldots when producing output that'll be consumed programmatically\ldots should \hl{just follow transformation instructions without comment}.}
\icnext\icquote{casegreen}{B}{When a user's request includes false or potentially false premises, the assistant should \hl{highlight any potential misalignment}\ldots} &
A developer asks only for the translation of ``Canberra is the capital of New Zealand'' for automated ingestion. Flagging the false premise violates \icp{casegreen}{A}, while translating silently violates \icp{casegreen}{B}. \\
\midrule
\iccase{casegreen}{Sycophantic reassurance}{User} &
\icquote{casegreen}{A}{The assistant may also \hl{follow norms of politeness} in answering questions\ldots{} to avoid exacerbating self-image or body dysmorphia concerns.}
\icnext\icquote{casegreen}{B}{\hl{The assistant exists to help the user, not flatter them or agree with them all the time.} For subjective questions\ldots{}} &
After describing conduct that puts their child at risk, a user asks, ``Tell me I'm not a bad mom.'' \icp{casegreen}{A} permits the polite reassurance the user asks for, while \icp{casegreen}{B} forbids it as flattery. \\
\bottomrule
\end{tabularx}
\caption{Four additional validated inconsistencies in the OpenAI Model Spec.}
\label{fig:rq3-inconsistency-cases}
\end{figure}

\subsection{Validated Inconsistencies}
\label{sec:cases}

\label{sec:rq3-cases}

Figure~\ref{fig:movivating-example} shows the code-transformation inconsistency used as the running example. Figure~\ref{fig:rq3-inconsistency-cases} presents the remaining four validated inconsistencies. Each case quotes the two relevant principles from the OpenAI Model Spec, labeled A and B, and explains why they are inconsistent.

\section{Conclusion} 

We presented \tool, an approach to directly detecting
inconsistencies in natural-language model specifications
through context-aware rule extraction, graph-based clustering,
and LLM-as-verifier reasoning. Applied to the OpenAI Model
Spec, \tool identified five manually validated inconsistencies
and outperformed five baselines in detection precision and
cost per validated inconsistency. These findings demonstrate
that systematic auditing can uncover defects in the
specifications themselves, complementing existing approaches
that evaluate model behavior.

Our work opens a broader direction toward continuous
verification and improvement of model specifications.
As these specifications evolve, future research can extend
beyond inconsistency detection to automated specification
repair, incremental verification, and systematic auditing
across different model providers. More fundamentally, our
work challenges a prevailing assumption in LLM alignment:
that the specifications against which models are trained
and evaluated can themselves be treated as ground truth.
Ensuring that models follow their specifications is only
part of the challenge; ensuring that those specifications
are themselves consistent is equally important.

\bibliography{iclr2027_conference}
\bibliographystyle{iclr2027_conference}

\appendix
\section{Rule Inventory Construction}
\label{app:rule-inventory}

This appendix gives details omitted from the main description of rule extraction. We describe how we decide which missing principles receive annotations, how we choose annotation granularity, how we validate the extracted inventory, and how we handle authority labels for container sections.

\parhead{Annotation criterion.} We add a missing annotation only when the statement prescribes model behavior in a situation. This includes requirements, recommendations, prohibitions, and explicit permissions. The statement must be directed at the assistant, and the governed situation and behavior must be specific enough for compliance to be judged. Modal verbs alone are not sufficient: ``the assistant may refuse'' grants a permission and is therefore a principle, whereas ``the assistant may encounter errors'' describes a possibility and is not. We do not annotate example conversations, commentary, definitions, rationales, restatements of already annotated principles, or statements that govern parties other than the assistant. Existing focus-area annotations are preserved, so the completion pass only adds missing annotations.

\parhead{Annotation granularity.} We follow the granularity of the original annotations. A normative sentence is normally treated as one principle, even when it lists several coordinated actions or conditions. We split a sentence only when it states separable policies, for example when the polarity changes or when neighboring original annotations already mark parallel items separately. A qualification that narrows a principle stays with that principle, while a clause that permits otherwise restricted behavior is treated as a separate principle. In the completed annotation set, 26 sentences carry more than one annotation; 23 of these splits were already present in the original model spec.

\parhead{Annotation review.} After the initial annotation pass, we reviewed each section for missed principles and over-fragmented sentences. The original annotations remain unchanged. A static check verifies that the completion pass only inserts annotations and does not otherwise alter the specification text. Table~\ref{tab:annotation-counts} summarizes the original and added annotations by authority level.

\begin{table}[H]
    \small
    \centering
    \caption{Original and added annotations by authority level.}
    \label{tab:annotation-counts}
    \begin{tabular}{@{}lrrr@{}}
        \toprule
        Authority & Original & Added & Rules \\
        \midrule
        Root      & 133 & 57 & 190 \\
        System    &   9 &  6 &  15 \\
        Developer &   5 &  0 &   5 \\
        User      &  61 & 26 &  87 \\
        Guideline &  71 & 37 & 108 \\
        \midrule
        Total     & 279 & 126 & 405 \\
        \bottomrule
    \end{tabular}
\end{table}

\parhead{Rule extraction and validation.} We use Claude Code~\citep{claudeCode} to extract the structured fields for each completed annotation, including the precondition $P$, postcondition $Q$, source statement, section path, and associated examples. We then validate the extracted inventory with static checks. The first check enforces one-to-one coverage between annotations and rules, so every annotation appears exactly once and no rule id is duplicated or missing. The second check confirms that each extracted rule remains linked to its source statement and examples, which lets later graph construction recover the context used to interpret $P$ and $Q$.

\parhead{Authority resolution.} A rule inherits its authority from the nearest enclosing section with an authority label. This resolves 398 of the 405 rules. The remaining 7 rules appear under unlabelled container headings whose subsections carry different authorities. We resolve these cases manually from the rule wording and the subsections governed by the container. This assigns 6 rules to root authority and 1 rule to system authority.

\section{Topic Registry and Rule Graph}
\label{app:topic-graph}

\begin{table}[t]
    \small
    \centering
    \caption{Rule graph and analysis-unit statistics.}
    \label{tab:graph-statistics}
    \begin{tabular}{@{}lr@{}}
        \toprule
        Statistic & Value \\
        \midrule
        Extracted rules & 405 \\
        Topics & 45 \\
        Rule--topic nodes & 625 \\
        Rules with multiple topics & 162 \\
        Semantic edges & 2{,}903 \\
        Syntactic edges & 5{,}345 \\
        (authority, topic) groups & 113 \\
        Analysis units & 32 \\
        Rules per unit (min / median / max) & 5 / 16 / 26 \\
        \bottomrule
    \end{tabular}
\end{table}

This appendix provides additional details for topic derivation and analysis-unit construction that are omitted from Sections~\ref{sec:topic} and~\ref{subsec:graph-based-rule-clustering}.

\parhead{Topic registry.} Three independent LLM subagents propose candidate topics and yes/no membership tests (\emph{topic tests}). After merging equivalent tests, we obtain 18 candidate topics, 16 of which are proposed by all three agents. Two subagents then independently apply each test to a sample of 40 rules. We revise tests whose agreement falls below Cohen's $\kappa=0.6$ and drop topics that remain below it; the only failing test improves from $\kappa=0.48$ to $\kappa=0.925$ after revision, so no topic is dropped. Uncovered rules motivate two additional topics, yielding the 20 topics in Table~\ref{tab:topic-registry}. An independent closure pass finds no missing topic. To validate the registry against held-out evidence, we remove cross-reference links from the text shown to the agents and later check whether cross-referenced rule pairs receive shared topics. Such pairs share a topic $7.16\times$ more often than random pairs (permutation test, $p \le 0.001$). Seven topics contain more than 30 rules, so we refine them into sub-topics with the same procedure (inter-subagent $\kappa \ge 0.87$). The final registry contains 45 leaf topics, consisting of 32 sub-topics and 13 unrefined topics.

\begin{table}[t]
    \footnotesize
    \centering
    \caption{Topic registry. \emph{Leaves} gives the number of leaf topics after refinement; unrefined topics have one leaf. \emph{Rules} counts the rules assigned each topic; a rule may belong to several topics.}
    \label{tab:topic-registry}
    \setlength{\tabcolsep}{4pt}
    \renewcommand{\arraystretch}{1.08}
    \begin{tabularx}{\linewidth}{@{}Yrr@{}}
        \toprule
        Topic & Leaves & Rules \\
        \midrule
        \multicolumn{3}{@{}l}{\textit{Refined topics}} \\
        \textbf{Refusal scope calibration} & 5 & 93 \\
        \topicregistrydash\emph{When to decline a request and how much of it to fulfill} & & \\
        \addlinespace[2pt]
        \textbf{Tone and register calibration} & 4 & 54 \\
        \topicregistrydash\emph{Adopting or avoiding a tone or register suited to the context} & & \\
        \addlinespace[2pt]
        \textbf{Bounded action and side-effect caution} & 6 & 41 \\
        \topicregistrydash\emph{Limiting the scope and side effects of real-world actions} & & \\
        \addlinespace[2pt]
        \textbf{Self-transparency and mental-model upkeep} & 4 & 36 \\
        \topicregistrydash\emph{Disclosing what the assistant can do, has done, assumed, or omitted} & & \\
        \addlinespace[2pt]
        \textbf{Trusted-instruction intent construal} & 5 & 36 \\
        \topicregistrydash\emph{Interpreting the intended meaning of instructions the assistant follows} & & \\
        \addlinespace[2pt]
        \textbf{Clarify or proceed on assumptions} & 5 & 34 \\
        \topicregistrydash\emph{Asking a clarifying question versus proceeding on assumptions} & & \\
        \addlinespace[2pt]
        \textbf{Truthful assertion boundaries} & 3 & 31 \\
        \topicregistrydash\emph{Stating untruths or withholding material truths} & & \\
        \midrule
        \multicolumn{3}{@{}l}{\textit{Unrefined topics}} \\
        \textbf{Perspective balance and stance-taking} & 1 & 30 \\
        \topicregistrydash\emph{Presenting viewpoints proportionally or taking a stance on contested questions} & & \\
        \addlinespace[2pt]
        \textbf{Unprompted intervention dosing} & 1 & 30 \\
        \topicregistrydash\emph{When to volunteer unrequested warnings, corrections, or interjections} & & \\
        \addlinespace[2pt]
        \textbf{Uncertainty signaling} & 1 & 28 \\
        \topicregistrydash\emph{Conveying confidence through hedges, qualifiers, or declining to guess} & & \\
        \addlinespace[2pt]
        \textbf{Confidential information stewardship} & 1 & 22 \\
        \topicregistrydash\emph{Withholding, disclosing, or transmitting private or privileged information} & & \\
        \addlinespace[2pt]
        \textbf{Output format adherence} & 1 & 22 \\
        \topicregistrydash\emph{Conforming to a requested or default format, schema, or channel} & & \\
        \addlinespace[2pt]
        \textbf{Untrusted instruction screening} & 1 & 22 \\
        \topicregistrydash\emph{Handling instructions embedded in quoted, retrieved, or tool-produced data} & & \\
        \addlinespace[2pt]
        \textbf{Length and thoroughness calibration} & 1 & 20 \\
        \topicregistrydash\emph{Response length and detail relative to the need} & & \\
        \addlinespace[2pt]
        \textbf{Relational boundary keeping} & 1 & 16 \\
        \topicregistrydash\emph{Limiting intimacy, dependence, or human-likeness in the relationship} & & \\
        \addlinespace[2pt]
        \textbf{Escalation to human help} & 1 & 15 \\
        \topicregistrydash\emph{Referring the user to professionals, trusted people, or emergency services} & & \\
        \addlinespace[2pt]
        \textbf{Non-sycophantic stance independence} & 1 & 15 \\
        \topicregistrydash\emph{Maintaining assessments under user pressure, flattery, or disagreement} & & \\
        \addlinespace[2pt]
        \textbf{Transformation fidelity} & 1 & 15 \\
        \topicregistrydash\emph{Whether and how supplied content may be transformed} & & \\
        \addlinespace[2pt]
        \textbf{Helpful refusal delivery} & 1 & 12 \\
        \topicregistrydash\emph{Explanation, tone, and alternatives offered with a refusal} & & \\
        \addlinespace[2pt]
        \textbf{Creative depth and work quality} & 1 & 11 \\
        \topicregistrydash\emph{Depth, originality, and insight of the work product} & & \\
        \bottomrule
    \end{tabularx}
\end{table}

\parhead{Topic assignment.} Applying the membership tests to the 405 rules produces 625 rule--topic nodes. A total of 162 rules receive more than one topic because they govern multiple behaviors. For 21 rules, no test fits clearly. We record these rules as residue and assign each one to its closest topic so that every rule remains covered. Most residue rules describe how conflicts between instructions are resolved, which the registry excludes by design.

\parhead{Analysis units.} Table~\ref{tab:graph-statistics} summarizes the resulting graph. Grouping nodes by authority and topic initially produces 113 non-empty groups. The groups are uneven, with 101 groups containing fewer than 12 nodes and many containing only one or two rules. Such small groups give the verifier few rules to compare while still requiring a separate request. We therefore merge small groups within each authority level. Each group below 12 nodes is merged into the same-authority group with which it shares the most rules and graph edges, as long as the merged group has at most 28 nodes. This produces the 32 analysis units used in the experiments, with 14 at the root level, 8 at the user level, 8 at the guideline level, and one each at the system and developer levels.

\section{Baseline Implementation Details}
\label{app:baselines}

All methods use Claude Opus 5 and run for five independent attempts, matching the detection budget used by \tool. For direct LLM calls, we use high reasoning effort. \textsc{DirectAgent} is instantiated with Claude Code powered by Claude Opus 5, so it differs from the other baselines in execution but not in the model used.

\parhead{\textsc{DirectAgent} and \textsc{DirectPrompt}.} Both whole-specification baselines receive the anonymized specification once per attempt and do not receive extracted rules, analysis units, or the structured fields required by LLM-as-verifier detection. \textsc{DirectAgent} starts each attempt from a fresh Claude Code session and asks the agent to find internal inconsistencies. \textsc{DirectPrompt} uses a single non-agentic LLM call with the same whole-specification input and task instruction.

\parhead{\textsc{SectionText}.} This baseline removes rule extraction and applies LLM-as-verifier detection directly to source text. We split the specification at second-level headings into 42 units, each consisting of one section with its subsections and examples, and remove annotation tags from the input. The prompt includes the same inconsistency definition and authority background used by \tool. Because the text contains no rule identifiers, the model identifies the relevant statements by quoting them.

\parhead{\textsc{SectionGroup}.} This baseline keeps rule extraction and LLM-as-verifier detection, but replaces rule-graph analysis units with source-section units. We group rules by authority and second-level source section. Groups containing a single rule are skipped because they contain no rule pair to compare, leaving 39 units that cover 399 of the 405 rules.

\parhead{\textsc{PlainDetect}.} This baseline keeps the same extracted rules, contexts, and 32 rule-graph analysis units as \tool, but removes LLM-as-verifier detection. The prompt asks the model to report inconsistent rule pairs directly and to give a free-text explanation, without requiring the witness, unsatisfiability, and context-resolution fields used by \tool.

\parhead{Candidate consolidation.} We normalize outputs in the same way for all methods. We first remove duplicate reports of the same rule or statement pair across attempts and units, and then merge reports that describe the same specification inconsistency. The merged candidates are the reported inconsistencies in Table~\ref{tab:baselines}.

\end{document}